\documentclass[twocolumn,showpacs,preprintnumbers,amsmath,prb,amssymb]{revtex4}

\usepackage{graphicx}
\usepackage{dcolumn}

\usepackage{bm}

\begin{document}
\bibliographystyle{aip}

\title{Raman studies of polycrystalline CaCu$_3$Ti$_4$O$_{12}$ under high-pressure}
\author{D. Valim, A.~G.~Souza Filho\footnote{Corresponding author, E-mail:
agsf@fisica.ufc.br}, P. T. C. Freire, A.P. Ayala, J. Mendes Filho}
\affiliation{Laborat\'orio de Espalhamento Raman, Departamento de
F\'{\i}sica, Universidade Federal do Cear\'a, 60455-900 Fortaleza,
Cear\'a, Brazil}
\author{A.F.L. Almeida, P.B.A. Fechine, A.S.B. Sombra} \affiliation{Laborat\'orio de
Telecomunica\c{c}\~oes e Ci\^encia e Enegenharia de Materiais -
LOCEM, Departamento de F\'{\i}sica, Universidade Federal do
Cear\'a, 60455-900 Fortaleza, Cear\'a, Brazil}

\date{\today}

\begin{abstract}
We report a Raman scattering study of polycrystalline
CaCu$_3$Ti$_4$O$_{12}$ (CCTO) under pressure up to 5.32\,GPa. The
pressure dependence of several Raman modes was investigated. No
anomalies have been observed on the phonon spectra thereby
indicating that the T$_{h}$ (Im$\bar 3$) structure remains stable
for pressures up to 5.32\,GPa. The pressure coefficients for the
observed modes were determined. This set of parameters was used
for evaluating the stress developed in CCTO thin films.
\end{abstract}
\maketitle


The complex perovskite CaCu$_3$Ti$_4$O$_{12}$ (CCTO) has been
recently reported as a material having the largest dielectric
constant (16000-18000) ever measured in the
laboratory.\cite{subramanian00,ramirez00,homes01} In addition to
this striking property, the dielectric constant is nearly constant
over a wide temperature range (from 100 to 600 K). Explaining the
anomalous dielectric constant has been an intriguing issue and
many models have been proposed. Some authors have attributed the
high values to extrinsic factors such as defects and grain
boundaries.\cite{cohen03,sinclair02} By performing first principle
calculations, the Vanderbilt's group has suggested that the source
of such giant dielectric constant should be related to extrinsic
effects since the contribution of the lattice effects was
estimated to be about 60.\cite{he02,he03} This value is extremely
smaller than what has been found in the experiments (10$^5$).
Recently, homogeneous CCTO thin films epitaxially grown were found
to exhibit a low-frequency permittivity of
$\approx$100.\cite{tselev03} This experimental result supports the
hypothesis that the giant dielectric constant obtained for bulk
CCTO comes from extrinsic effects. Owing to these remarkable
properties, CCTO is being considered as a very promising material
for application in microelectronics, mainly in capacitive
elements. With the miniaturization of the devices and having in
view the large and prompt integration with the current technology,
CCTO thin films are the most likely systems to be used in
CCTO-based devices. Such systems have been investigated and
several reports on growth,\cite{lin02,fang03,jha03}
dielectrics,\cite{tselev03,zhao031} dc and ac electrical
resistivity,\cite{chen03} and optical
properties\cite{litvinchuk03} have already been reported.

Thin films are expected to exhibit built-in stress due to the
grain boundary, reduced dimensionality, and the lattice mismatch
with the various substrate used for growing them. The residual
stress affects the mechanical, electrical and optical properties
of the films. Therefore, it is important to investigate the stress
developed in the films in order to improve the growth methods and
to tailor desired physical properties. Raman spectroscopy has been
established as a powerful and versatile technique
(non-destructive, contact-less and non-invasive) for investigating
the stress in films by monitoring the phonon frequency shifts of
the films relative to their bulk counterparts. For carrying out
such analysis, the knowledge of the pressure dependence of the
Raman spectra of polycrystalline CCTO can be the basis for probing
and interpreting built-in stress in CCTO films.

So far, the Raman modes of CCTO have been measured in
ceramics\cite{almeida02}, single
crystals,\cite{kolev02,koitzsch02}, and thin
films.\cite{litvinchuk03}. These investigations have been devoted
to study the polarization and temperature dependence of the Raman
spectra.\cite{kolev02,koitzsch02} In this work we report the
effects of pressure on the Raman spectra of polycrystalline CCTO.
We found that the mode frequencies exhibit a linear behavior with
pressure. No anomalies have been observed from ambient pressure up
to 5.32 GPa thus indicating the absence of structural
instabilities (phase transitions) in this pressure range. The
modes are sensitive to the applied pressure. The pressure
coefficients $\alpha=\partial \omega/\partial P$ were determined
and these values will be useful for calculating the Gr\"uneisen
parameters. Finally, we used our set of pressure coefficients
along with the Raman data available in the literature
\cite{litvinchuk03} for evaluating the stress in CCTO thin films.

\bigskip

Commercial oxides Ca(OH)$_2$ (Vetec, 97\% with 3\% of CaCO$_3$),
CaCO$_3$ (Aldrich, 99\%), TiO$_2$ (Aldrich, 99\%), and CuO
(Aldrich, 99\%) were used for preparing CCTO ceramics. The CCTO
samples were prepared by the conventional powder-sintering
technique. The starting materials were weighted according to the
stoichiometric ratios and mixed thoroughly in an agate mortar. The
mixed powder was calcined at 900$^{o}$C for 12 h, and the
resulting samples were sintered in air at 1050$^o$C for 24 h.
High-pressure Raman experiments were performed using a diamond
anvil cell (DAC) with 4:1 methanol : ethanol mixture as the
transmitting fluid. The pressure calibration was achieved by using
the well known pressure shift of the ruby luminescence lines. The
pressure dependent Raman spectra were obtained with a
triple-grating spectrometer (Jobin Ivon T64000) equipped with a
N$_2$ cooled charge coupled device (CCD) detection system. The
line 514.5 nm of an argon ion laser was used as excitation. An
Olympus microscope lens with a focal distance f$=20.5$ nm and
numeric aperture N.A.$=0.35$ was employed to focus the laser beam
on the sample surface.

\bigskip

\begin{figure}
{\includegraphics[angle=0,height=3.5in]{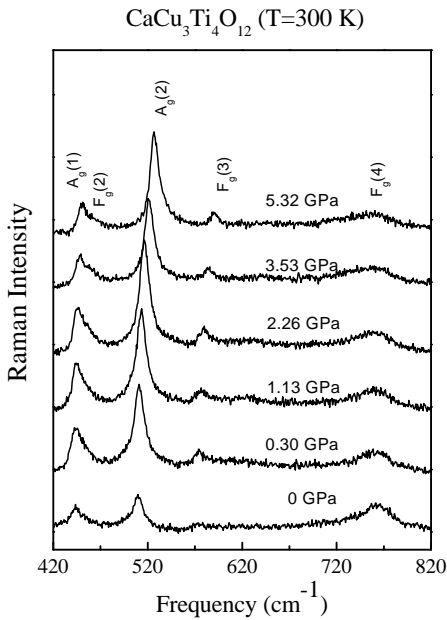}}
\caption{Pressure dependent Raman spectra of CCTO.}\label{fig1}
\end{figure}

Before discussing our results we briefly summarize some properties
of CCTO. This material has a body-center cubic primitive cell
containing 20 atoms and belonging to the centrosymmetric
T$_h$(Im$\bar{3}$) group. Standard group theory analysis predicts
that the Raman active modes are distributed among the irreducible
representation as 2A$_g$+2E$_g$+4F$_g$.\cite{kolev02}  The lower
trace in Fig.\,\ref{fig1} shows the Raman spectrum of CCTO at
ambient pressure and room temperature. CCTO is a weak scatter and
only five of the eigth predicted modes are observed at around 444,
453, 510, 576 and 761 cm$^{-1}$. Based on lattice dynamics
studies\cite{he02,he03} and polarized Raman measurements,
\cite{kolev02} the mode symmetries are identified as
A$_{g}$(1)(444 cm$^{-1}$), F$_{g}$(2)(453 cm$^{-1}$),
A$_{g}$(2)(510 cm$^{-1}$), F$_{g}$(3)(576 cm$^{-1}$) and
F$_{g}$(4)(761 cm$^{-1}$). Following lattice dynamics calculations
the 444, 453, and 510 cm$^{-1}$ are TiO$_6$ rotationlike modes.
The peak at 576\,cm$^{-1}$ is assigned to the Ti-O-Ti
anti-stretching mode of the TiO$_6$ octahedra.\cite{kolev02} The
F$_g$(4) mode has been predicted to be observed at about 710
cm$^{-1}$. We have assigned the mode 761 cm$^{-1}$ to the
symmetric stretching breathing of the TiO$_{6}$ octahedra. Its low
intensity is typical of the structures containing shared units
where the neighbors octahedra underdamp the symmetric vibrations.
First-principle calculations\cite{he02} predict this breathing
mode to be observed at 739 cm$^{-1}$. This value is much more
close to 761 cm$^{-1}$ we have observed than the prediction of
classical lattice dynamics.\cite{kolev02}

\begin{figure}
{\includegraphics[angle=0,height=3.5in]{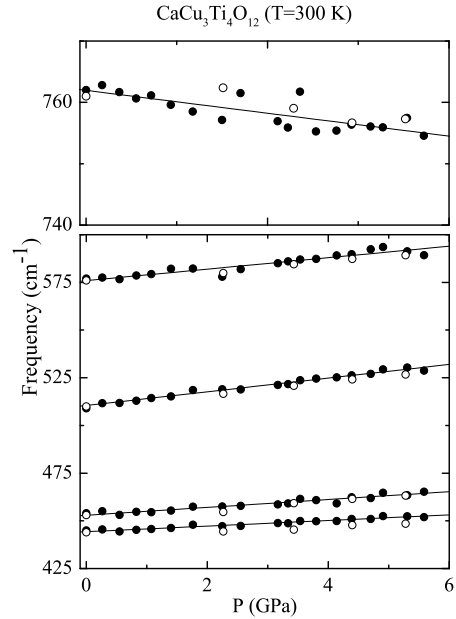}}
\caption{Frequency vs Pressure plots for CCTO Raman modes. The
solid and open circles stand for compression and decompression
runs. The solid lines are fit to the experimental data using
$\omega(P)$=$\omega_{0}+\alpha P$.}\label{fig2}
\end{figure}

Upon increasing pressure, it is evident from the spectra (see
Fig.1), that the material remains in its initial configuration up
to the maximum pressure we have reached in our experiments. In
order to discuss in detail the pressure effects on CCTO phonon
frequencies we have constructed the frequency ($\omega$) vs.
pressure (P) plots. The results for both compression (solid
circles) and decompression (open circles) experiments are shown in
Fig.\,\ref{fig2}. All peaks exhibit a linear behavior
$\omega(P)$=$\omega_{0}+\alpha P$ and both frequency intercepts
($\omega_{0}$) and pressure coefficients ($\alpha$) are listed in
Table\,\ref{t002}. Note, that all modes are sensitive to pressure
and large $\alpha$ values were observed. For the breathing mode
$\partial\omega/\partial P<$0, indicating a possible structural
instability that might lead to a structural phase transition at
higher pressures. Finally, the pressure-dependent data also reveal
that the linewidths change slightly with pressure. This result
suggests that the pressure does not induce any disorder in CCTO.

\begin{table*}
\caption{Frequency intercepts $\omega_0$ and pressure coefficients
$\alpha$ for CCTO.} \label{t002}
\begin{ruledtabular}
\begin{tabular}{ccc}
Mode   & $\omega_0$ (cm$^{-1}$)& $\alpha$ (cm$^{-1}$GPa$^{-1}$) \\
\hline
A$_g$(1)     & 444  & 1.47   \\
F$_g$(2)      & 453  & 2.06   \\
A$_g$(2)     & 510  & 3.60   \\
F$_{g}$(3)   & 576  & 3.00   \\
F$_{g}$(4)   & 761  & -1.20  \\
\end{tabular}
\end{ruledtabular}
\end{table*}

Provided the pressure coefficients we have the ground knowledge
for discussing the Raman scattering results in CCTO thin films.
Litvinchuk et al.\cite{litvinchuk03} report the Raman scattering
studies of CCTO thin films prepared by pulsed laser deposition on
a (001) LaAlO$_3$ substrate. When compared with bulk CCTO, these
authors have observed the Raman modes in the film are shifted
toward higher frequency by an amount of 4-7\,cm$^{-1}$. Similarly
to spectra of bulk CCTO, the A$_g$(2) mode is the most intense
mode measured in CCTO films. It was observed at 517\,cm$^{-1}$
which is 7\,cm$^{-1}$ higher than in CCTO bulk. We attribute this
upshift to the stress developed during the growth process. The
pressure behavior of the A$_g$(2) mode has been determined as
$\omega (P)=510+3.6P$, being $\omega$ and $P$ in units of
cm$^{-1}$ and GPa, respectively. By using this equation and the
value of $\omega$ measured in the CCTO film we calculate that the
stress developed in the CCTO film reported in
ref.\,\cite{litvinchuk03} is 1.94$\pm$0.3\,GPa.

\bigskip

Summarizing, we studied polycrystalline CCTO under high pressure.
No evidence for pressure-induced phase transition was found in the
0-5.32\,GPa pressure range. The frequencies of all Raman modes
exhibit a linear dependence on pressure. We also determined the
pressure coefficients $\alpha=\partial \omega/\partial P$ for all
modes. This set of parameters was used for evaluating the built-in
stress in CCTO thin films prepared by pulsed laser deposition. The
$\alpha$ values would be also useful for calculating the
Gr\"uneisen parameters for CCTO as the Young´s modulus become
available. Finally, this work improves the characterization data
of CCTO that will be useful for feeding back theoretical models
and allowed to improve the understanding of the Raman spectra
properties of CCTO thin films.

\bigskip
D.V. and A.G.S.F. acknowledge financial support from  the
Brazilian agencies CNPq and CAPES (PRODOC grant No. 22001018),
respectively. The authors acknowledge Dr. I. Guedes for a critical
reading of the manuscript and partial support from Brazilian
agencies CNPq, FUNCAP and FAPESP.


\end{document}